\documentclass[twocolumn,prd,aps]{revtex4-1}
\usepackage{graphicx}
\usepackage{hyperref}
\usepackage{amssymb}
\usepackage{amsmath}
\usepackage{xcolor}
\usepackage{lineno}
\usepackage{enumitem}

\newcommand{\jcap}{{J. Cosmol. Astropart. Phys.}}
\newcommand{\mnras}{{Mon. Not. Roy. Astron. Soc.}}
\newcommand{\aj}{{Astron. J.}}
\newcommand{\aap}{{Astron. Astrophys.}}
\newcommand{\apjl}{{Astrophys. J. Lett.}}
\newcommand{\physrep}{{Phys. Rep.}}

\begin{document}

\title{Ultra-High-Energy Cosmic Rays from Tidally-Ignited White Dwarfs}

\author{Rafael \surname{Alves Batista}$^1$}
\email{rafael.alvesbatista@physics.ox.ac.uk}

\author{Joseph Silk$^{1,2,3,4}$}
\email{joseph.silk@physics.ox.ac.uk}

\affiliation{
$^1$ Department of Physics - Astrophysics, University of Oxford, DWB, Keble Road, OX1 3RH, Oxford, UK\\
$^2$ Institut d’Astrophysique  de Paris,  98 bis bd Arago, F-75014, Paris, France\\
$^3$ The Johns Hopkins University, Department of Physics \& Astronomy, 3400 N. Charles St., Baltimore, MD 21218, USA\\
$^4$ AIM-Paris-Saclay, CEA/DSM/IRFU, CNRS, Univ Paris 7, F-91191, Gif-sur-Yvette, France}

\begin{abstract}
Ultra-high-energy cosmic rays (UHECRs) can be accelerated by tidal disruption events of stars by black holes. We suggest a novel mechanism for UHECR acceleration wherein white dwarfs (WDs) are tidally compressed by intermediate-mass black holes (IMBHs), leading to their ignition and subsequent explosion as a supernova. Cosmic rays accelerated by the supernova may receive an energy boost when crossing the accretion-powered jet. The rate of encounters between WDs and IMBHs can be relatively high, as the number of IMBHs may be substantially augmented once account is taken of their likely presence in dwarf galaxies. Here we show that this kind of tidal disruption event naturally provides an intermediate composition for the observed UHECRs, and suggest that dwarf galaxies and globular clusters are suitable sites for particle acceleration to ultra-high energies.
\end{abstract}



\maketitle

\section{Introduction}
\label{sec:intro}

An open question in astrophysics is related to the origin and acceleration mechanisms of the ultra-high-energy cosmic rays (UHECRs). Their extreme energy ($E \gtrsim 10^{18}\;\text{eV}$) poses a problem for simple models of particle acceleration. Measurements by the Pierre Auger Observatory favor a mixed composition at energies $E \gtrsim 10^{18.6} \; \text{eV}$, which becomes increasingly heavier with energy~\citep{auger2014a,auger2014b}. The recent combined spectrum-composition fit by the Auger Collaboration also suggests a mixed composition at the sources~\citep{auger2017a}, although the results are model-dependent.

At energies below $\sim \, 0.1 \; \text{EeV}$ ($1 \; \text{EeV} \equiv 10^{18} \; \text{eV}$), the cosmic-ray spectrum can be described by the usual supernova paradigm: supernovae accelerate CR protons up to energies $\sim 10^{15} \; \text{eV}$; this leads to the so-called \emph{knee} of the cosmic-ray spectrum. Heavier nuclei of atomic number $Z$ can be accelerated up $E_{\text{max}} \sim Z \times 10^{15} \; \text{eV}$; for iron ($Z=26$) this would imply a feature in the spectrum coinciding with the \emph{second knee}. CRs of energies above the so-called \emph{ankle} ($E \sim 4 \times 10^{18} \; \text{eV}$) are believed to be mostly of extragalactic origin. Betwen the second knee and the ankle there is a clear need for a local (possibly galactic) component~\cite{hillas2005a,thoudam2016a}, whose acceleration mechanism remains to be found. Extensions of the usual acceleration mechanism by supernovae have been suggested to explain this component (see e.g. Refs.~\cite{voelk1988a,bell2001a}).

A difficulty common to many UHECR acceleration models concerns the efficiency at which a source accelerates heavy nuclei. Gamma-ray bursts (GRBs), for example, are inefficient, as opposed to unipolar inductors operating in pulsars and magnetars. Outflows of active galactic nuclei are also seemingly suitable accelerator sites, although the abundance of nuclei such as iron is low.

Tidal disruption events (TDEs) by supermassive black holes (SMBHs) have been suggested as the underlying mechanism for particle acceleration to ultra-high energies~\citep{farrar2009a}, and invoked to explain anisotropy results by the Pierre Auger Observatory and Telescope Array (TA)~\citep{pfeffer2017a}. 

In this paper we focus on a different kind of TDE in which a white dwarf (WD) is disrupted and ignited by an intermediate-mass black hole (IMBH).

IMBHs are hypothetical objects whose existence have not been unambiguously confirmed. They provide a natural link between stellar-mass and supermassive black holes. They are believed to be found in dwarf galaxies, with an occupation fraction for active IMBHs of order 1\% in both x-ray~\citep{miller2015a,baldassare17} and optical surveys~\citep{moran2014a}, as well as in globular clusters~\citep{kiziltan2017a,perera2017a,oka2017a}. A substantial number of AGNs in dwarf galaxies may be missed because of soft x-ray absorption~\citep{chen2017a}. 
Many of the dwarf ``problems'', ranging from frequency, absence of cusps, too-big-to-fail and baryon fraction, can be accounted for by feedback from IMBH  in the early gas-rich phase of dwarf galaxy evolution~\citep{silk17a}.
We infer that quiescent IMBHs are likely to occur in at least 10\% of observed dwarfs as well as in  massive globular clusters, and moreover are plausibly expected to be produced in  many of the failed dwarfs predicted by galaxy formation theory in the context of the $\Lambda$CDM model, if only as relic building blocks of SMBHs~\citep{rashkov2014a}.

The typical IMBH mass lies in the range between those of stellar mass and supermassive black holes, i.e.,  $\sim 10^2 - 10^5 \rm M_\odot $. WDs are the end product of main-sequence stars with  masses up to  $\sim 8 \ \rm M_\odot .$ 
Due to the abundance of WDs, in dense environments they are prone to collide and/or merge with other objects. In some cases, WDs can be disrupted by IMBHs and explode as supernovae. The observation of the gamma-ray burst GRB060218 and its associated supernova, SN2006aj~\citep{campana2006a}, has been interpreted by \citet{shcherbakov2013a} as such an encounter. A similar interpretation for the object Swift J1644+57 has also been suggested by \citet{krolik2011a}.

\section{The model}
\label{sec:model}

If a star of mass $M_\star$ is close to a black hole of mass $M_{bh}$, tidal forces may deform the star, stretching it along the orbital direction and squeezing it in the perpendicular direction. The net effect of this deformation is an increase in pressure which may be enough to trigger nuclear burn, causing the explosion of the star as a low-luminosity supernova~\citep{carter1982a}. As suggested by \citet{luminet1989a,luminet1989b}, this can happen for white dwarfs with masses between $0.07 M_\odot$ and $0.50 M_\odot$ that are not able to normally burn helium but which can be ignited by tidal disruptions. Similarly, for stars of $M_\star \gtrsim 0.5 M_\odot$, TDEs can trigger the burning of carbon and oxygen, of  which  white dwarfs in this mass range are largely composed.

Stars can be tidally disrupted by BHs  if their distance  is
\begin{equation}
r_{t} \lesssim 0.1 M_{bh, 3}^{1/3} M_\star^{-1/3} R_{\star, -2} \; R_\odot
\end{equation} 
where $R_{\star, -2}$ is the stellar radius in units of $10^{-2}R_{\odot}$, $M_\star$ is the mass of the star in solar masses, and $M_{bh, 3}$ is the BH mass in units of $1000 M_\odot$. Let $r_p$ be the pericenter distance between the star and the black hole. The encounter strength ($\beta$) is usually defined as the ratio between the tidal radius $r_t$ and the pericenter distance $r_p$. The Schwarzschild radius of the black hole is $R_g = 2GM_{bh}/c^2$, thus constraining the maximum mass of a black hole for which the star can be tidally disrupted, to~\citep{rosswog2009a}
\begin{equation}
M_{bh,max} \approx 1.2 \times 10^{5} R_{\star,-2}^{3/2} M_{\star}^{-1/2} \; M_\odot.
\end{equation} 
Note that SMBHs with masses $M_{bh} \gtrsim 10^6 M_\odot$ cannot tidally disrupt white dwarfs.

Once ignited, the star will unavoidably become a supernova. The relative abundances of elements synthesized by white dwarfs in tidally-induced supernova events will be used to fix the initial composition of UHECRs in our model. Naturally, these abundances depend upon the initial composition of the star. Low-mass WDs ($0.07M_\odot \lesssim M_\star \lesssim 0.5M_\odot$) are mainly composed of helium, whereas those with masses $M_\star \gtrsim 0.5M_\odot$ have as their primary composition intermediate-mass nuclei such as carbon and oxygen. We now introduce two simple models, with the abundances ($a_i$) of each species obtained from estimates by~\citet{rosswog2009a}\\
{\it model I}: WDs with masses $M_\star < 0.5M_\odot$ are composed initially of helium, with abundances $a_\text{He} = 0.861$, $a_\text{C} = 0.12$, $a_\text{Si} = 0.11$, and $a_\text{Fe} = 0.016$;\\
{\it model II}: WDs with masses $M_\star > 0.5M_\odot$ are composed initially of equal amounts of carbon and oxygen, with abundances $a_\text{He} = 0.568$, $a_\text{C} = 0.044$, $a_\text{O} = 0.096$, $a_\text{Ne} = 0.002$, $a_\text{Mg} = 0.002$, $a_\text{Si} = 0.114$, and $a_\text{Fe} = 0.173$. \\ 
Model I is for a helium WD of $M_\star = 0.2M_\odot$ disrupted by a BH of $M_{bh} = 10^3M_\odot$, whereas model II is for a carbon/oxygen WD of $M_{\star} = 1.2M_\odot$ encountering a black hole of $M_{bh} = 500M_\odot$. The mass of the black hole and white dwarf, as well as the proximity of the encounter, are expected to affect the relative abundances of each element, so that models I and II are intended to be representative cases rather than an exact model.

We now want to estimate the luminosity in UHECRs for such encounters. For this, we follow \citet{evans2015a} and \citet{shcherbakov2013a}.

Debris of the explosion of the star will be accreted onto the BH at a rate
\begin{equation}
	\label{eq:accretionRate}
	\dot{M}_{bh} \equiv \frac{dM_{bh}}{dt} = \frac{1}{3} \frac{M_\star}{t_{fb}} \left( \frac{t}{t_{fb}} \right)^{-\frac{5}{3}},
\end{equation}
where $t_{fb}$ is the characteristic time for bound material to fall into the black hole, given by
\begin{equation}
	\label{eq:fbTime}
	t_{fb} \approx  100 \alpha^{-3} R_{\star, -2}^{3/2} M_{bh,3}^{1/2} M_\star^{-1} \; \text{s},
\end{equation}
where $\alpha \equiv \beta \sqrt{\xi} / \kappa$, with $\kappa \sim 0.5$ being a model-dependent parameter, and $\xi \equiv \Delta R_\star / R_\star$ designating the deformation factor for a spread $\Delta R_\star$.  In WD-ignition scenarios,  $\beta$ is typically $\sim 10$. Simulations find $\xi \sim 4$~\citep{evans2015a}. Therefore, our benchmark value is $\alpha \sim 40$.

Combining equations \ref{eq:accretionRate} and \ref{eq:fbTime}, we can write $\dot{M}_{max} \equiv M_\star/(3t_{fb})$, which reads
\begin{equation}
	\dot{M}_{max} \approx 10^4  \alpha^{3} R_{\star,-2}^{-3/2} M_\star^2 M_{bh,3}^{-1/2} \; M_\odot \, \text{yr}^{-1}.
	\label{eq:Mdotmax}
\end{equation}
Note that the accretion rate decays with $t^{-5/3}$. This is believed to be a universal property of tidal disruption events~\citep{rees1988a,evans1989a,phinney1989a,laguna1993a,rosswog2009a}. 

As the material from the supernova explosion falls into the black hole, part of the magnetic field of the former star is advected toward the event horizon, increasing the total magnetic energy in the vicinity of the black hole, yielding 
\begin{equation}
	B_{bh} \sim 4 \times 10^{11} \dot{M}_{bh,4}^{1/2} M_{bh,3}^{-1} \; \text{G},
	\label{eq:Bbh}
\end{equation}
where $\dot{M}_{bh,4} \equiv \dot{M}_{bh} \times 10^{4} \; M_\odot \, \text{yr}^{-1}$.
The magnetic field $B_{bh}$ is composed of a turbulent random component arising from magnetorotational instability~\citep{balbus1991a,hawley1991a,balbus1998a}, and the regular (poloidal) component generated via dynamo amplification of seed fields ~\citep{brandenburg2005a}. These fields can reach values of $B_{bh} \sim 10^{12}\;\text{G}$~\citep{brandenburg2005a}, which can be maintained during a short period of time. 

In Eq.~\ref{eq:fbTime},  there is an $\alpha^3$ dependence. In studies by \citet{stone2013a} and \citet{guillochon2013a}, $\alpha$ has been found to be independent of the encounter strength $\beta$. These differences, however, do not invalidate our results. They would cause a change of one order of magnitude in the estimates of Eq.~\ref{eq:Bbh} for $\beta \sim 10$, and a factor of a few for $\beta \lesssim 5$.

We now assume that this is a Kerr black hole and that energy can be extracted from the BH via a mechanism such as~\citet{blandford1982a} or \citet{blandford1977a}. For details, the reader can refer to e.g.~\citep{blandford1977a,rees1982a,mckinney2004a}. The jet is launched due to the presence of a poloidal magnetic field, as shown in general relativistic magnetohydrodynamical simulations performed  by \citet{tchekhovskoy2011a} and \citet{mckinney2012a}. 

The main difference between this and other tidal disruption models found in the literature is the ignition of the WD. About 10\% of the material of the supernova explosion is accreted onto the BH, powering the jet. The remaining material contains cosmic rays accelerated by the supernova. When these cosmic rays cross the jet, they can receive an energy boost of $\sim \gamma^2$, as proposed by \citet{caprioli2015a} for the case of AGN jets accelerating cosmic rays produced within their host galaxies; here  $\gamma \approx \Psi^{-1}$ is the Lorentz factor of the jet, and $\Psi$ its opening angle. Because the accelerated cosmic rays are not part of the accretion flow, in this scenario,  acceleration would not occur near the source, but rather in optically thin regions away from the base of the jet. Consequently, photodisintegration is not expected to be relevant, implying that the primary composition of the UHECRs remains virtually unaltered. The jet itself may also accelerate cosmic rays to ultra-high energies, but in this case interactions with the envelope could heavily degrade the energy of the escaping particles. Nevertheless, this is important to determine the neutrino and gamma-ray signatures of the relevant  processes.

\section{Validating the model}

It is well-known that supernovae can accelerate cosmic rays up to energies of at least $\sim Z \times 10^{15} \; \text{eV}$ via diffusive shock acceleration (DSA). Features in the cosmic-ray spectrum, namely the second knee and the ankle, provide a very strong case for the existence of a local (possibly galactic) component of CRs that cannot be explained with simple linear DSA theory~\cite{hillas2005a,thoudam2016a}, thus requiring an extension of the DSA or possibly a completely new acceleration mechanism.  Although the exact details of such processes are not relevant for our phenomenological description, there are interesting possibilities that naturally extend the spectrum of cosmic rays accelerated by supernovae beyond the second knee that should be mentioned. One was suggested by Bell \& Lucek~\cite{bell2001a} (see also Refs.~\cite{lucek2000a,bell2004a}) and consists of the non-linear amplification of magnetic fields due to the backreaction of the streaming CRs on the medium.  Other mechanisms predict acceleration in winds of companion objects~\cite{voelk1988a,crocker2015a}.
The mechanisms commonly found in the literature, including the aforementioned ones, predict $E_{max} \sim Z \times 10^{17} \; \text{eV}$. 

Following the supernova explosion, there are two groups of CRs that are accelerated. The first one is composed by galactic CRs present in the interstellar medium (hereafter referred to as \emph{ISM composition}), and the second relates to the abundances of nuclei synthesized in the explosion (designated as \emph{SN composition}). While for $E \lesssim Z \times 10^{15} \; \text{eV}$,  the acceleration of ISM-composition CRs via DSA (or a similar mechanism)  is virtually guaranteed, the same is not true for the case of SN composition as the synthesized nuclei would not be accelerated by forward shocks, but rather by reverse shocks, whose efficiency is about 10\% the efficiency of the former. 
The viability of particle acceleration in reverse shocks has been demonstrated in many studies~\cite{ellison2005a,zirakashvili2010a,telezhinsky2012a,lee2014a}. Because DSA of CRs with ISM composition is not enough to explain the composition of galactic cosmic rays, an additional mechanism such as reverse shocks is required, whereby non-ISM nuclei can be produced and subsequently accelerated~\cite{parizot2014a}. As a matter of fact, particle acceleration in reverse shocks has been proposed in Refs.~\cite{ramaty1996a,lingenfelter1998a} to explain a long-standing problem related to the abundance of elements such as lithium, beryllium, and boron in galactic cosmic rays. While this is rather speculative, it is not unreasonable to expect a contribution from nuclei synthesized by a supernova. 
From the theoretical viewpoint, particle acceleration could also take place in black-hole winds produced via a mechanism similar to a generalization of advection-dominated accretion flows~\cite{blandford1999a,tombesi2015a}.

Now the main ingredients of our model are available: cosmic-ray nuclei are synthesized in the SN and subsequently accelerated to energies of at least $E_{max} \sim \text{a few} \times 10^{17} \; \text{eV}$; an accretion-powered jet is launched by the BH following the tidal disruption event. We now invoke a one-shot mechanism inspired by Ref.~\cite{caprioli2015a}, whereby CRs crossing the jet receive an energy boost proportional to $\Gamma^2$, with $\Gamma \sim 10$ being the Lorentz factor of the jet. A slightly higher $E_{max}$ than predicted by linear DSA theory and/or a jet Lorentz factor of $\Gamma \approx 15$ are sufficient to explain the existence of CRs at energies $\sim 10^{20} \; \text{eV}$. 

Not all IMBH-WD disruptions will lead to supernovae, but up to $\sim 10\%$ of them will. The Pierre Auger Observatory  has derived lower bounds on the density of sources based on the absence of clustering in the arrival directions of UHECRs; the least stringent bound is $n_0 \gtrsim 6 \times 10^{-6} \; \text{Mpc}^{-3}$ and the most stringent $n_0 \gtrsim 7 \times 10^{-4} \; \text{Mpc}^{-3}$, at 95\% confidence level, depending on how sources are distributed and on the strength of extragalactic magnetic fields~\citep{auger2013a}. 
Under the simple assumption that the rate of IMBH-WD encounters is constant over time, we obtain a true density $n \sim 10^{-1} \; \text{Mpc}^{-3}$ averaged over a Hubble time, using estimates from Ref.~\cite{baumgardt2004a}. Due to deflections in intervening galactic and extragalactic magnetic fields, the apparent rate of these events would be $\dot{n}_{s} \sim n / \delta t$~\cite{waxman1996a,takami2012a}, where $\delta t$ is the typical spread in arrival times, given by~\cite{waxman1996a}
\begin{equation}
	\delta t \simeq 10^{7} B_\text{nG}^2 \ell_\text{Mpc} E_{20}^{-2} \;\, \text{yr},
	\label{eq:tdelay}
\end{equation}
and where $B_\text{nG} = B / \text{nG}$ is the magnetic field in nG, $\ell_\text{Mpc}$ is the coherence length in Mpc, and $E_{20}$ is the cosmic-ray energy in units of $10^{20} \; \text{eV}$.
This equation holds for sources closer than $D \lesssim 75 \; \text{Mpc}$, which corresponds to the interaction horizon at ultra-high energies. Within a sphere of radius 75 Mpc, most of the volume is likely occupied by voids (for a detailed discussion see e.g. Ref.~\cite{alvesbatista2017a}), and we conservatively estimate $\dot{n}_s \sim 10^{-8} \; \text{Mpc}^{-3}\,\text{yr}^{-1}$ using the upper limit for magnetic fields in voids, $B \sim 1 \; \text{nG}$, assuming a coherence length $\ell \simeq 1 \; \text{Mpc}$~\cite{planck2016a}.


For a typical IMBH of 1000$M_\odot$, and a WD of $1 M_{\odot}$ and $10^{-2} R_\odot$, the maximum accretion rate is given by Eq.~\ref{eq:Mdotmax}. The energy injection rate ($\dot{E}$) is 
\begin{equation}
	\dot{E} = \dot{n}_s E_{CR} = \frac{n}{\delta t} \epsilon E_{tot} ,
\end{equation}
with $\epsilon$ being the efficiency of conversion of the total energy of the progenitor star ($E_{CR} \equiv \epsilon E_{tot}$) into cosmic rays. The total energy, $E_{tot}$, can be as high as $\sim 10^{51} \; \text{erg}$. Estimates by \citet{katz2009a} suggest $\dot{E} \sim 10^{44} \; \text{erg} \, \text{Mpc}^{-3} \, \text{yr}^{-1}$, so that in order to satisfy energy requirements, one would need $\epsilon / \delta t \gtrsim 10^{-6} \; \text{yr}^{-1}$. We estimate that $\epsilon \ll 10\%$. Recalling that $\delta t \propto B^2 \ell$ (from Eq.~\ref{eq:tdelay}), a simple decrease in the magnetic field by less than one order of magnitude would already render our  model viable. 

While we could have derived the efficiency factor from considerations about the mass function of black holes, rates of IMBH-WD TDEs, etc, the uncertainties related to these quantities and the likely dependence of $\epsilon$ on other parameters justify our approach.

Using the abundances of the elements that compose the WD, in addition to the elements synthesized in the supernova, we can fix the UHECR composition. As a proxy for the composition, we use the depth of the maximum of the air shower ($\langle X_{max} \rangle$). Therefore, the only free parameters of the model are the spectral index of the injected cosmic rays ($\alpha$), the maximum rigidity to which they are accelerated ($\mathcal{R}_{max}$), and the source evolution, assumed here to follow the star formation rate. This is a major source of uncertainty in our model, as the source evolution is a combination of the evolution of WDs, which follow the star formation rate, and of IMBHs, which is not well known~\cite{volonteri2011a}.

In order to illustrate the predictions made by the model, we compute the expected spectrum and composition by simulating the propagation of UHECRs from the sources to Earth.  Assuming that the sources are randomly distributed in the comoving volume, we simulate the propagation of UHECRs using the CRPropa 3 code~\citep{alvesbatista2016a}. We consider all relevant interactions and energy loss processes, namely: photopion production, pair production, photodisintegration, nuclear decay, and adiabatic losses due to the expansion of the universe. The main target for interactions are the cosmic microwave background (CMB) and the extragalactic background light (EBL). In particular, we use the EBL model by \citet{gilmore2012a}. The simulations use as input the fraction of each nuclear species following. The spectrum has the form
\begin{equation}
	\dfrac{dN_{inj}}{dE} = \sum\limits_{i=1}^{26} J_0  a_i   E_{inj}^{-\alpha} f(E_{inj}, Z_i),
	\label{eq:dNdE}
\end{equation}
where $a_i$ is the relative abundance of each nucleus of atomic number $Z$, $J_0$ is the overall normalization factor, and the function $f(E_{inj})$ is 1 for $E_{inj}<Z \mathcal{R}_{max}$ and $\exp(1 - E_{inj}/(Z \mathcal{R}_{max}))$ otherwise. 

Note that forward and reverse shocks may have different maximal rigidities ($\mathcal{R}_{max}$), spectral indices ($\gamma$), and injected elemental abundances ($a_i$). Moreover, these parameters may be different for the two source populations (helium and carbon-oxygen WDs). Therefore, the actual spectrum is a combination of Eq.~\ref{eq:dNdE} for forward and reverse shocks, as well as for both source populations.

It is worth stressing that the combined spectrum-composition fit~\cite{auger2017a} of the data collected by the Pierre Auger Observatory suggests hard spectral indices, which is consistent with a contribution of reverse shocks to the total flux, as forward shocks tend to have softer spectral indices compared to reverse shocks~\cite{ptuskin2013a}. 

\begin{figure*}[ht]
	\centering
	\includegraphics[width=.495\textwidth]{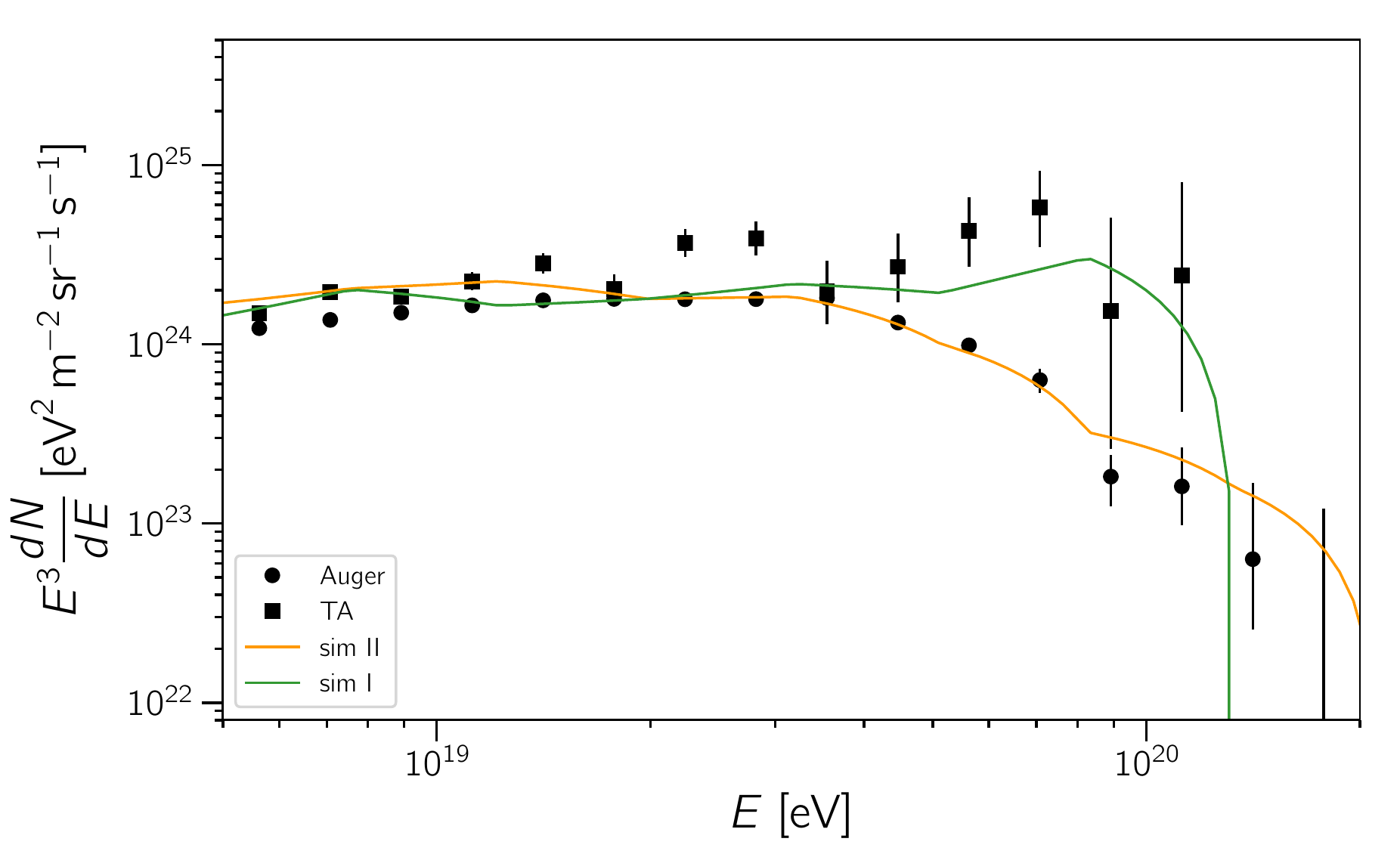}
	\includegraphics[width=.495\textwidth]{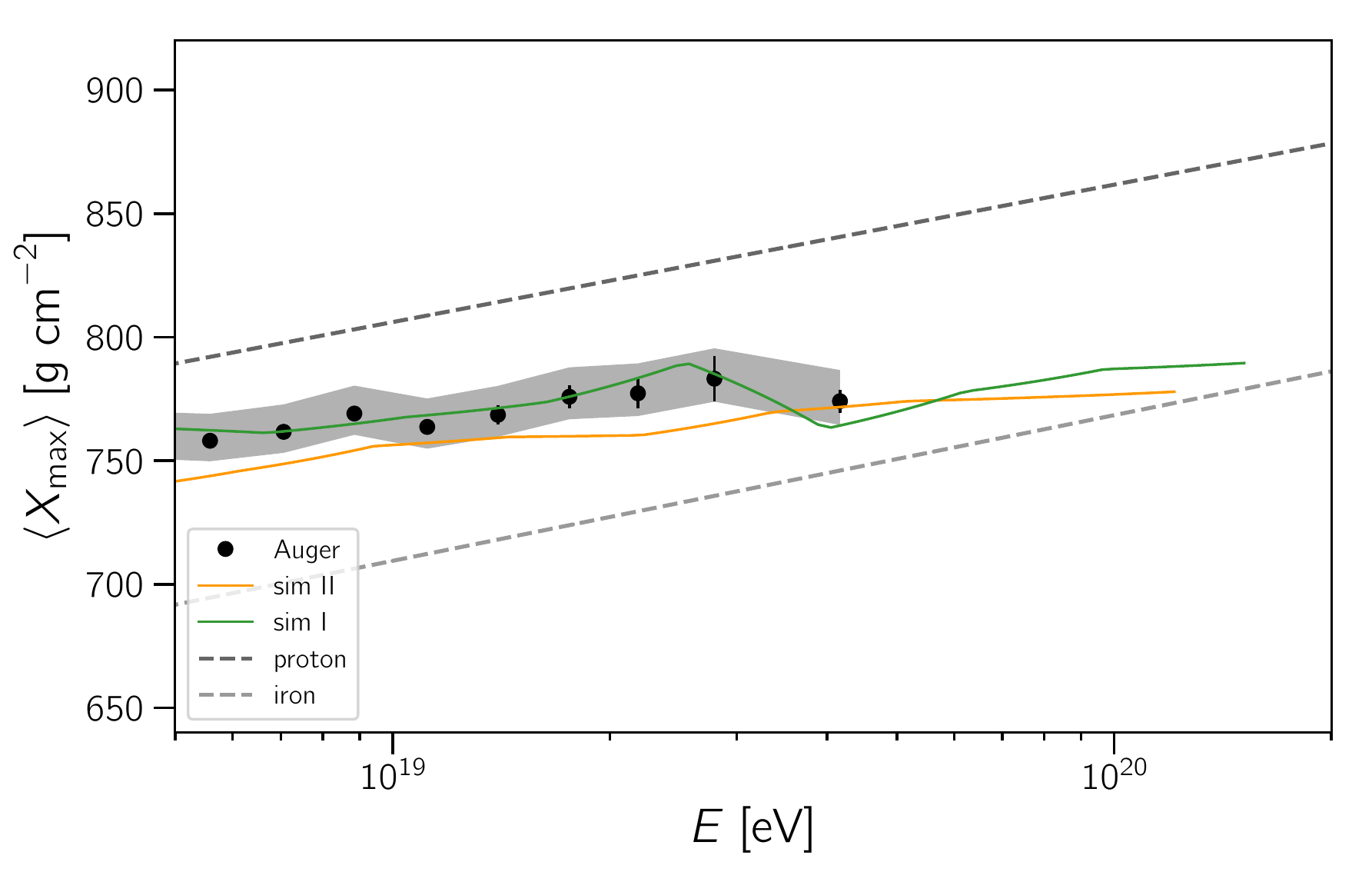}
	\caption{Spectrum (left), $\langle X_{max} \rangle$ (right panel), for models I and II. These plots are for the case of $\mathcal{R}_{max} = 8 \; \text{EV}$, assuming that sources evolve following the star formation rate. Spectral indices are $\alpha = 1.2$ (sim II) and $\alpha = 0.4$ (sim I). The spectrum is arbitrarily normalized with respect to Auger data at $E = 10 \; \text{EeV}$. Hatched gray bands represent systematic uncertainties.}
	\label{fig:speccomp}
\end{figure*}

The results are shown in Fig.~\ref{fig:speccomp} for both models. In order to obtain $\langle X_{max} \rangle$, we have used the parametrization given by \citet{auger2013b}, assuming the hadronic interaction model EPOS-LHC ~\citep{pierog2015a} for the development of cosmic-ray showers in the atmosphere. 

Note that Fig.~\ref{fig:speccomp} is merely for illustration purposes. In reality, $\mathcal{R}_{max}$ and $\alpha$ depend on the acceleration model, being the maximal attainable energy limited by the magnetic field capable of confining cosmic rays within the source environment. The spectrum and composition should be simultaneously constrained as done by the~\citet{auger2017a}. Nevertheless, due to the many uncertainties involved in such a combined fit, whose evaluation is beyond the scope of this work, we simply present a couple of scenarios to qualitatively discuss the model.  Other combinations of $\alpha$, $\mathcal{R}_{max}$, and source evolution may lead to better agreement with the measurements. Our model was devised to hold for $E \gtrsim 10^{18.6} \; \text{eV}$, since this is the energy at which the galactic and extragalactic components of the cosmic-ray spectrum overlap, according to some models~\citep{deligny2014a}.

These predictions are subject to uncertainties related both to the propagation of UHECRs from their sources to Earth and to the relative abundance of each element synthesized in the supernova. The latter has been discussed by \citet{rosswog2009a} and is inherent to the modeling of the explosion. The former depends on parameters such as the exact distribution of sources, photonuclear cross sections, and the EBL model adopted, as shown in Ref.~\cite{alvesbatista2015a}. Moreover, magnetic fields may also play a role in the spectral shape, although to which extent it is uncertain. 

Gravitational waves produced by these kinds of events are unlikely to be detectable  by Advanced LIGO, since they  would be at the experimental sensitivity limit, amplitude $\sim 10^{-22}$ and frequency $\sim 10 \; \text{Hz}$)~\citep{ligo2016a}. 

A detailed calculation of the electromagnetic signatures of IMBH-WD tidal disruption events can be found in \citet{macleod2014a}, and suggests signatures in x-rays.

\section{Discussion and Conclusions}

We have presented an original model for UHECR acceleration in IMBH-WD TDEs with ignition of the white dwarf,  that naturally explains the intermediate-mass composition of UHECRs observed by Auger.
We have described the acceleration of cosmic rays to ultra-high energies by assuming a one-shot reacceleration mechanism in jets generated via TDEs of WDs by IMBHs. Cosmic rays are first accelerated by the supernova and subsequently they are reaccelerated by the accretion-powered jet. A number of mechanisms for cosmic-ray acceleration by supernovae and reacceleration by jets could be invoked. Nevertheless, one should keep in mind that in order for intermediate/heavy nuclei to escape, the source environment has to be optically thin.

Promising environments to search for IMBH-WD systems are globular clusters. Globular clusters are abundant in the universe and even in the Milky Way~\citep{rashkov2014a}. About 10\% of all stars disrupted in globular clusters are WDs. Moreover, N-body simulations by \citet{baumgardt2004a} suggest that one in every ten of these clusters contain IMBHs. Recent observations confirm that nearby massive globular clusters may contain IMBH~\citep{kiziltan2017a,perera2017a,oka2017a}.

Other interesting candidates are dwarf galaxies, as they may host IMBHs that are possibly required to seed the formation of supermassive black holes~\citep{volonteri2003a,moran2014a}. The large number of relic dwarfs expected in CDM-motivated galaxy formation models provides a potentially rich source of IMBHs that have been proposed to play an important role in determining the Milky Way galaxy's baryon fraction~\citep{peirani2012a}. Moreover  the inefficient formation of the central SMBH by merging IMBHs should leave a significant population of surviving IMBHs~\citep{islam03a}. The Milky Way may consequently harbor IMBHs in regions where WDs are fairly abundant. Therefore, one can expect a few IMBH-WD encounters within our galaxy contributing to the observed UHECR spectrum. 

A possible tracer of IMBH-WD tidal disruption events are ultra-luminous x-ray sources (ULXs), whose luminosities ($\gtrsim 10^{39} \; \text{erg}\,\text{s}^{-1}$) exceed those of any stellar process known. \citet{shcherbakov2013a} has interpreted the x-ray flare from GRB060218 as such an event. In particular, the observation of an associated supernova (SN2006aj) strengthens this interpretation. 

Detailed numerical simulations of IMBH-WD encounters could be used to improve the estimates of the composition of the exploding star, thus fixing the composition of UHECRs in the neighborhood of sources. Electromagnetic, neutrino, and gravitational-wave counterparts of such events could provide a way to test acceleration models in IMBH-WD systems. Nevertheless, because tidal ignitions of WDs by IMBHs are transient events, we do not expect to see UHECRs in coincidence with photons, neutrinos, and gravitational waves, since time delays would be incurred by intervening magnetic fields. In general because of the short time-scales and higher luminosities of white dwarf TDEs~\citep{krolik2012a}, they are expected to be prominent in flux-limited transient surveys~\citep{baumgardt2006a}.

\acknowledgments

RAB acknowledges support from the John Templeton Foundation. The work of JS has been supported in part by ERC Project No. 267117 (DARK) hosted by the Pierre and Marie Curie University-Paris VI, Sorbonne Universities and CEA-Saclay.

\bibliographystyle{apsrev4-1}
\bibliography{references}

\end{document}